\documentclass[reqno]{article}%12pt%
\usepackage{alltt}
\usepackage{graphicx}
\title{``Field-shell" of the self-interacting quantum electron}
\author{Peter Leifer$^{1,2}$, Taha Massalha$^{2}$}
\date{$^{1}$Technion - Israel Institute of Technology, Department of Education in Technology and Science, Haifa, Israel;\\
$^{2}$The Academic Arab College for Education, Physics Department, Haifa, Israel \\
leifer@bezeqint.net, tahamas@gmail.com }
\begin{document}
\maketitle
\begin{abstract}
Self-interacting dynamics of non-local Dirac's electron has been proposed. This dynamics was revealed by the projective representation of operators corresponding to spin/charge degrees of freedom. Energy-momentum field is described by the system of quasi-linear ``field-shell" PDE's following from the conservation law expressed by the affine parallel transport in $CP(3)$ \cite{Le1}. We discuss here solutions of these equations in the connection with the following problems: curvature of $CP(3)$ as a potential source of electromagnetic fields and the self-consistent problem of the electron mass.
\end{abstract}
\vskip 0.1cm
\section{Introduction}
Primarily there were two mathematical approaches to the formulation of quantum theory. The first one (developed by Hiesenberg) makes accent on the non-commutative character of new ``quantized" dynamical variables whereas the second one (developed by Schr\"odinger) replaces ordinary differential Hamilton's equations of classical dynamics by linear differential equations in partial derivatives associated with Hamilton-Jacobi equation \cite{Schr1}. Both approaches are equivalent in the framework of so-called optics-mechanics analogy and comprise of the fundament of modern quantum mechanics. This analogy, however, is limited by itself in very clear reasons: mechanics is merely a coarse approximation (even being generalized to many-dimension dynamics of Hertz) and the ``optics" of the action waves is too tiny for description of complicated structure of ``elementary" quantum particles. It was realized already under the first attempts to synthesize relativistic and quantum principles.

Analysis of the foundations of quantum theory and relativity shows that there are
relativity of two types. One of them is the symmetries relative space-time transformations of whole quantum setup reflects, say, the \emph{first order of relativity}. Different type of (state-dependent) symmetries is realized in the state space of ``quantum particles" relative local infinitesimal variation of flexible quantum setup (\emph{second order of relativity or ``super-relativity"} \cite{Le1,Le2,Le3}). Gauge invariance is a particular case of this type symmetry. Analysis shows too that it is impossible to use ordinary primordial elements like particles, material points, etc., trying to build consistent theory. Even space-time cannot conserve its independent and a priori structure. Therefore the unification of relativity and quantum principles may be formalized if one uses new primordial elements: pure quantum degrees of freedom and the classification of their motions. So, the rays of quantum states will be used instead of material points (particles) and complex projective Hilbert state space $CP(N-1)$ where these states move under the action of unitary group $SU(N)$ instead of space-time \cite{Le2,Le3}.

Then:

1. Dynamical variables are in fact the generators of the group of symmetry and their non-commutative character is only a consequence of the curvature of the group manifold \cite{G}. State-dependent realization of $SU(N)$ generators as vector fields on $CP(N-1)$ evidently reveals the non-trivial global geometry of $SU(N)$ and it coset sub-manifold \cite{Le2,Le4}.

2. Attempts ``to return" in the Minkowsky space-time (after second quantization) from the Schr\"odinger's configuration space is successful for statistical aims but they are contradictable on the fundamental level of single quantum particle (which without any doubt does exist!) and therefore should be revised. In fact initially one should \emph{delete global space-time} by transition to ``co-moving frame" and after virtual infinitesimal displacement of generalized coherent state (GCS) of electron \emph{to restore state-dependent local dynamical space-time}.

3.  The physically correct transition from quantum to classical mechanics arose as a serious problem immediately after the formulation of ``wave mechanics" of Schr\"odinger \cite{Schr2}. The failure to build stable wave packet for single electron from solutions of linear PDE's lead to statistical interpretation of the wave function. The further progress in the theory of non-linear PDE's like  sin-Gordon or KdV renewed generally the old belief in possibility to return to deterministic quantum physics of ``elementary" particles \cite{Rajaraman}. Unfortunately, the main technical results concern non-linear PDE's are given by classical models. Even pure quantum field models were frequently reduced to well known classical non-linear PDE's. However, the fundamental problem is to find ``first" physical principles capable derive quantum non-linear PDE's. In fact one should invert the Schr\"odinger's original approach: from quantum wave equation one should come to non-linear field equations and to get classical dynamical equations as a reasonable approximation. We will use Dirac's wave (non-secondly quantized) equation for self-interacting electron since the problem of ``free" electron is contradictable and requires further clarification. It is used here only for the first orientation.

The revision mentioned above (see point (2)) proposed here intended to derive new non-linear quantum equations for self-interacting non-local electron.
Notice, that new field equations could not contain \emph{arbitrary potential} as it was in the case of Schr\"odinger or Dirac equations. This potential should be generated by the spin/charge self-interaction. One of the consistent way is to use quasi-linear field PDE's following from conservation law that has been already discussed \cite{Le1,Le2,Le3,Le4}. It is provided by state-dependent local non-Abelian ``chiral" gauge field acting on $CP(3)$ as a tangent vector fields.

Perturbation of generalized coherent state of $G=SU(4)$ of the electron is studied in the vicinity of the stationary degenerated state given by ordinary (not secondly quantized) Dirac's equation. This perturbation is generated by coset transformations $G/H=SU(4)/S[U(1) \times U(3)]=CP(3)$ as an analog of the infinitesimal Foldy-Wouthuysen transformations \cite{Le1}. Self-interaction arose due to the curvature of the projective Hilbert space $CP(3)$ and the state-dependent dynamical space-time (DST) is built during ``objective quantum measurement" \cite{Le3}.

The physical sense of the electron's model proposed here is that the electron
is \emph{a cyclic motion of quantum degrees of freedom} in projective Hilbert state
space $CP(3)$. Namely, it is assumed that the motion of spin/charge degrees of freedom
comprises of stable attractor in the state space, whereas its ``field-shell" in
dynamical space-time arises as a consequence of the local conservation law of
energy-momentum vector field.

\section{Eigen-dynamics and local dynamical variables}
The standard QM tells us
what is the spectrum of dynamical variables, but it is silent about dynamics
of morphogenesis of stationary quantum states (since they are states of motion).
The quantum mechanics assumes the priority of the Hamiltonian given
by some classical model which henceforth should be ``quantized". It
is known that this procedure is ambiguous. In order to avoid the
ambiguity, we intend to use a {\it quantum state} itself and the
invariant conditions of its conservation and perturbation. These
invariant conditions are rooted into the global geometry of the
dynamical group manifold. Namely, the geometry of $G=SU(N)$, the
isotropy group $H=U(1)\times U(N-1)$ of the pure quantum state, and
the coset $G/H=SU(N)/S[U(1)\times U(N-1)]$ geometry, play an essential
role in the quantum state evolution \cite{Le4}. The stationary states
i.e. the states of motion with the least action may be
treated as {\it initial conditions} for GCS evolution. Particulary
they may represent a local minimum of energy (local vacuum).

It is well known
that even in the general case of non-bilinear function $\bar{a}(\psi,\psi^*)$ giving
eigen-values as minimums at eigen-vector is defined in fact on the complex
projective Hilbert space $CP(N-1)$ \cite{Weinberg}.
The local coordinates
\begin{equation}
\pi^i_{(j)}=\cases{\frac{\psi^i}{\psi^j},&if $ 1 \leq i < j$ \cr
\frac{\psi^{i+1}}{\psi^j}&if $j \leq i < N-1$}
\end{equation}\label{1}
for ray solution of the eigen-problem have been used
instead of the vector solution with additional freedom
of a complex scale multiplication \cite{Le3}.

Now we will introduce the local dynamical variables (LDV's) correspond
to the internal $SU(N)$ group symmetry and its breakdown. They should be
expressed now in terms of the local coordinates $\pi^k$. Thereby they
will live in geometry of $CP(N-1)$ with the Fubini-Study metric
\begin{equation}
G_{ik^*} = [(1+ \sum |\pi^s|^2) \delta_{ik}- \pi^{i^*} \pi^k](1+
\sum |\pi^s|^2)^{-2}
\end{equation}\label{2}
and the affine connection
\begin{eqnarray}
\Gamma^i_{mn} = \frac{1}{2}G^{ip^*} (\frac{\partial
G_{mp^*}}{\partial \pi^n} + \frac{\partial G_{p^*n}}{\partial
\pi^m}) = -  \frac{\delta^i_m \pi^{n^*} + \delta^i_n \pi^{m^*}}{1+
\sum |\pi^s|^2}.
\end{eqnarray}\label{3}
 Hence the internal dynamical
variables and their norms should be state-dependent, i.e. local in
the state space \cite{Le4}. These local dynamical variables realize
a non-linear representation of the unitary global $SU(N)$ group in
the Hilbert state space $C^N$. Namely, $N^2-1$ generators of $G =
SU(N)$ may be divided in accordance with the Cartan decomposition:
$[B,B] \in H, [B,H] \in B, [B,B] \in H$. The $(N-1)^2$ generators
\begin{eqnarray}
\Phi_h^i \frac{\partial}{\partial \pi^i}+c.c. \in H,\quad 1 \le h
\le (N-1)^2
\end{eqnarray}\label{4}
of the isotropy group $H = U(1)\times U(N-1)$ of the ray (Cartan
sub-algebra) and $2(N-1)$ generators
\begin{eqnarray}
\Phi_b^i \frac{\partial}{\partial \pi^i} + c.c. \in B, \quad 1 \le b
\le 2(N-1)
\end{eqnarray}\label{5}
are the coset $G/H = SU(N)/S[U(1) \times U(N-1)]$ generators
realizing the breakdown of the $G = SU(N)$ symmetry of the GCS.
Furthermore, the $(N-1)^2$ generators of the Cartan sub-algebra may
be divided into the two sets of operators: $1 \le c \le N-1$ ($N-1$
is the rank of $Alg SU(N)$) Abelian operators, and $1 \le q \le
(N-1)(N-2)$ non-Abelian operators corresponding to the
non-commutative part of the Cartan sub-algebra of the isotropy
(gauge) group. Here $\Phi^i_{\sigma}, \quad 1 \le \sigma \le N^2-1 $
are the coefficient functions of the generators of the non-linear
$SU(N)$ realization. They give the infinitesimal shift of the
$i$-component of the coherent state driven by the $\sigma$-component
of the unitary  field $\exp(i\epsilon \lambda_{\sigma})$ rotating by the
generators of $Alg SU(N)$ and they are defined as follows:
\begin{equation}
\Phi_{\sigma}^i = \lim_{\epsilon \to 0} \epsilon^{-1}
\biggl\{\frac{[\exp(i\epsilon \lambda_{\sigma})]_m^i \psi^m}{[\exp(i
\epsilon \lambda_{\sigma})]_m^j \psi^m }-\frac{\psi^i}{\psi^j} \biggr\}=
\lim_{\epsilon \to 0} \epsilon^{-1} \{ \pi^i(\epsilon
\lambda_{\sigma}) -\pi^i \},
\end{equation}\label{6}
\cite{Le2,Le4}. The partial derivatives are defined here as usual: $\frac{\partial }{\partial \pi^i} = \frac{1}{2}
(\frac{\partial }{\partial \Re{\pi^i}} - i \frac{\partial }{\partial
\Im{\pi^i}})$ and $\frac{\partial }{\partial \pi^{*i}} = \frac{1}{2}
(\frac{\partial }{\partial \Re{\pi^i}} + i \frac{\partial }{\partial
\Im{\pi^i}})$.

\section{The affine gauge fields in $CP(3)$}
Since the system of eigen-vectors belonging to degenerated eigenvalue is defined up to unitary transformations, the approximate calculation of eigenvalues and corresponding eigen-state vectors in the conditions of degeneration is the natural place for application of the unitary group geometry. For example, the solution of the problem of small denominators arising in the framework of perturbation theory is based in fact on the geometry of $CP(1)$, see for example \cite{Fock}.
But there is a different interesting application of this geometry.

Pseudo-electric and pseudo-magnetic fields arose as gauge fields with singular potentials at the degeneration points of Hamiltonian spectrum \cite{Berry1}.
However, it is well known that the spectrum structure of degenerated linear operator is unstable relative small perturbation. Besides this it is known that re-parametrization of the monopole problem leads to its singularity-free Lagrangian form \cite{Aitchison}. Therefore such degenerated Hamiltonians cannot serve as a source of real electromagnetic potentials.

In order to understand the true source
of the electromagnetic fields we would like to study the affine unitary gauge fields arose under breakdown (reconstruction) of global $G=SU(4)$ symmetry of degenerated bi-spinors states of quantum electron to the local gauge group $H=S[U(1)\times U(3)]$ acting by state-dependent generators on ``phase space" $CP(3)$. Deformation of quantum state under the action of the geodesic flow in $CP(3)$ is treated here as
process of the quantum motion.

We will work with Dirac's operator of energy-momentum
\begin{equation}\label{7}
 \hat{\gamma}^{\mu}p_{\mu} =i\hbar \hat{\gamma}^{\mu}\frac{\partial}{\partial x^{\mu}}
\end{equation}
instead of the Hamiltonian. This combined operator acts in the direct product $S=C^4 \times H_D$, where $H_D$ means a Hilbert space of differentiable functions.
Such splitting seems to be artificial and we try to find more flexible construction
of energy-momentum operator. Namely, more reasonable to work in the fibre bundle and only in a section to have locally the splitting into ``external" and ``internal" degrees of freedom.

Lets apply to this operator the similarity transformation (transition in ``moving frame" freezing the action of the differentiation in space-time coordinates) with help the canonical unitary operator. In the case of pseudo-euclidian coordinates $x^{\mu}$ it is possible to use simply the plane wave $U_{gauge}=exp(-\frac{i}{\hbar}P_{\mu}x^{\mu})$.

The state of free Dirac's electron is the plane wave ``modulated" by the bi-spinor
\begin{eqnarray}\label{8}
|\Psi(x)>=\left(                                                               \begin{array}{cc}
\psi_1  \\
\psi_2 \\
\psi_3 \\
\psi_4 \\
\end{array}
\right) \exp{\frac{-i}{\hbar}P_{\mu}x^{\mu}}.
\end{eqnarray}
This state may be expressed in local coordinates
as follows: for $a=1$ one has
\begin{eqnarray}\label{9}
\psi^1(\pi^1_{j(p)},\pi^{2}_{j(p)},\pi^{3}_{j(p)})=e^{i\alpha_1}(1+
\sum_{s=1}^{3}|\pi^s_{j(p)}|^2)^{-1/2}
\end{eqnarray}
and for $a: 2\leq a = i \leq 4$ one has
\begin{eqnarray}\label{10}
\psi^i(\pi^1_{j(p)},\pi^{2}_{j(p)},\pi^{3}_{j(p)})= e^{i\alpha_1}\pi^i_{j(p)}(1+
\sum_{s=1}^{3}|\pi^s_{j(p)}|^2)^{-1/2}.
\end{eqnarray}
Notice, the local projective coordinates of free Dirac's electron states do not contain space-time degrees of freedom. Indeed: there are scale-invariant dimensionless local projective coordinates $(\pi^1, \pi^2, \pi^3)$ of free
electron in $CP(3)$, i.e. with relative components of bi-spinos of
stationary state derived from ordinary homogeneous system of eigen-problem
\begin{eqnarray}
mc^2 \psi_1+c(p_x-ip_y)\psi_4+cp_z \psi_3=E\psi_1 \cr
mc^2 \psi_2+c(p_x+ip_y)\psi_3-cp_z \psi_4=E\psi_2 \cr
-mc^2 \psi_3+c(p_x-ip_y)\psi_2+cp_z \psi_1=E\psi_3 \cr
-mc^2 \psi_4+c(p_x+ip_y)\psi_1-cp_z \psi_2=E\psi_4 .
\end{eqnarray}\label{11}
It is easy to see \cite{Le3} that transition from the system of homogeneous
equations to reduced system of non-homogeneous equations for rays has single-value
solution in each map for local coordinates $(\pi^1, \pi^2, \pi^3)$. It is possible only if determinant of the reduced system
$D=(E^2-m^2c^4-c^2p^2)^2 \neq 0$. Say, in the map $U_1:\{\psi_1 \neq 0\}$,
for $E=\sqrt{m^2c^4+c^2p^2}$
one has
\begin{eqnarray}\label{12}
\pi^1=0, \quad
\pi^2=\frac{cp_z}{mc^2+E} \quad
\pi^3=\frac{c(p_x+ip_y)}{mc^2+E}.
\end{eqnarray}
It is naturally to use these scale-invariant functional
variables in order to establish relation between spin-charge degrees of freedom
and energy-momentum distribution of electron in dynamical space-time (DST) since
the off-shell condition $D=(E^2-m^2c^4-c^2p^2)^2 \neq 0$ opens the way
for its self-interacting. New dispersion law will be established
due to formulation of conservation law of energy-momentum.

We need to restore  space-time degrees of freedom in state-dependent dynamical space-time for self-interacting electron that presumably should generate surrounding electromagnetic field.
In order to clarify this process it is useful to refer to the Berry's formula for 2-form  \cite{Berry1}.  Being applied to state vector $|\Psi(x)>$ in the local coordinates $\pi^i$, one has antisymmetric second-rank tensor
\begin{eqnarray}\label{13}
V_{ik*}(\pi^i)&=& \Im \sum_{a=1}^{4} \{\frac{\partial
\psi^{a*}}{\partial \pi^i} \frac{\partial \psi^a}{\partial \pi^{k*}} -
\frac{\partial \psi^{a*}}{\partial \pi^{k*}} \frac{\partial
\psi^a}{\partial \pi^i} \} \cr &=& -\Im [(1+ \sum |\pi^s|^2) \delta_{ik}-
\pi^{i^*} \pi^k](1+ \sum |\pi^s|^2)^{-2}= - \Im G_{ik^*}
\label{form}.
\end{eqnarray}
This is simply the imaginary part of the Fubini-Study quantum metric
tensor. There are following important differences between
original Berry's formula referring to arbitrary parameters and this
2-form in local coordinates inherently related to eigen-problem.

1. The $V_{ik*}(\pi^i) = i G_{ik^*}$ is the singular-free expression.

2. It does not contain two eigen-values, say, $E_n, E_m$ explicitly,
but implicitly $V_{ik*} = i G_{ik^*}$ depends locally on the choice of single
$\lambda_p$ through the dependence in local coordinates
$\pi^i_{j(p)}$. Even in the case of degenerated eigen-value,
the reason of the anholonomy lurks in the curvature of $CP(3)$ and therefore
it has intrinsically invariant and stable character.

3. It is impossible of course directly identify $V_{ik*} = i G_{ik^*}$ with electromagnetic tensor
$ F_{ij}=A_{j,i}-A_{i,j}$. We try to understand how the geometry of $CP(3)$ generates electromagnetic potentials in terms of ``filed-shell" equations for energy-momentum
\cite{Le1}.

Following changes have been done in order to get these equations:

1. The local projective coordinates $(\pi^1, \pi^2, \pi^3)$ of generalized coherent state (GCS) of electron will be use instead of Berry parameters $\mathbf{X}$ of a Hamiltonian $H(\mathbf{X})$  and therefore the quantum metric tensor is the Fubini-Study metric tensor of in $CP(3)$ \cite{Le1,Le2};

2. The iteration procedure of the Foldy-Wouthuysen coset transformations was be replaced by the infinitesimal action of local dynamical variables (LDV) represented by tangent vector fields on $CP(3)$ diffeomorphic to the coset sub-manifold
$G/H = SU(4)/S[U(1)\times U(3)$;

3. Affine parallel transport of the energy-momentum vector field on $CP(3)$ agrees with Fubini-Study metric will be used instead of ``adiabatic renormalization" \cite{Berry1} of the Dirac operator.

\section{ ``Field-shell" equations for non-local quantum electron}
The ``field-shell" equations arose as a consequence of the conservation law of energy-momentum vector field \cite{Le1,Le2,Le3}. They have some ``lump" solutions which should be carefully studied. In particular it is clear that quantum nature of derived field quasi-linear PDE's without any references to classical analogy could shed the light on the their generic connection with Hamilton-Jacobi classical equations and de Broglie-Schr\"odinger optics-mechanics analogy.

Quantum lump of non-local electron should presumably serve as extended source of electromagnetic field. This lump may be mapped onto dynamical space-time if one assumes that transition from one GCS of the electron to another is accompanied by dynamical transition from one Lorentz frame to another.
Thereby, infinitesimal Lorentz transformations define small
``dynamical space-time'' coordinates variations. It is convenient to take
Lorentz transformations in the following form
\begin{eqnarray}\label{14}
ct'&=&ct+(\vec{x} \vec{a}_Q) \delta \tau \cr
\vec{x'}&=&\vec{x}+ct\vec{a}_Q \delta \tau
+(\vec{\omega}_Q \times \vec{x}) \delta \tau
\end{eqnarray}
where we put for the parameters of quantum acceleration and rotation the definitions $\vec{a}_Q=(a_1/c,a_2/c,a_3/c), \quad
\vec{\omega}_Q=(\omega_1,\omega_2,\omega_3)$ \cite{G} in order to have
for $\tau$ the physical dimension of time. The expression for the
``4-velocity" $ V^{\mu}$ is as follows
\begin{equation}\label{15}
V^{\mu}_Q=\frac{\delta x^{\mu}}{\delta \tau} = (\vec{x} \vec{a}_Q,
ct\vec{a}_Q  +\vec{\omega}_Q \times \vec{x}) .
\end{equation}
The coordinates $x^\mu$ of imaging point in dynamical space-time serve here merely for the parametrization of the energy-momentum distribution in the ``field
shell'' arising under ``morphogenesis" described by quasi-linear field
equations \cite{Le1,Le3}. Notice, since we discarded pointwise particles, the energy-momentum of electron should be represented by some distribution in DST.

The conservation law of the energy-momentum vector field in $CP(3)$ during evolution will be expressed by the equation of the affine parallel transport
\begin{equation}\label{16}
\frac{\delta P^i}{\delta \tau} = \frac{\delta (P^{\mu}\Phi_{\mu}^i(\gamma_{\mu}))}{\delta \tau} = 0,
\end{equation}
which is equivalent to the following system of four coupled quasi-linear PDE for dynamical space-time distribution of energy-momentum ``field shell" of quantum state and ordinary differential equations for relative amplitudes
\begin{equation}\label{17}
V^{\mu}_Q (\frac{\partial
P^{\nu}}{\partial x^{\mu} } + \Gamma^{\nu}_{\mu \lambda}P^{\lambda})= -\frac{c}{\hbar}(\Gamma^m_{mn} \Phi_{\mu}^n(\gamma) + \frac{\partial
\Phi_{\mu}^n (\gamma)}{\partial \pi^n}) P^{\nu}P^{\mu},
\quad \frac{d\pi^k}{d\tau}= \frac{c}{\hbar}\Phi_{\mu}^k P^{\mu},
\end{equation}
which is in fact the \emph{equations of characteristic} for linear ``super-Dirac"
equation
\begin{equation}\label{18}
i P^{\mu}\Phi_{\mu}^i(\gamma_{\mu})\frac{\partial \Psi}{\partial \pi^i} =mc \Psi
\end{equation}
that supposes ODE for single ``total state function"
\begin{equation}\label{19}
i \hbar \frac{d \Psi}{d \tau} =mc^2 \Psi
\end{equation}
with the solution for variable mass $m(\tau)$
\begin{equation}\label{20}
\Psi(T) = \Psi(0)e^{-i\frac{c^2 }{\hbar}\int_0^T m(\tau) d\tau}.
\end{equation}
In this article we will discuss only the ``field-shell" equations (11).

\section{Solutions of ``field-shell" equations and dispersion laws for self-interacting electron}
We will discuss now the solution of the ``field-shell" equations (39).
The theory of these equations is well known. Particularly, our system is
the system with identical principle part $V^{\mu}_Q$ which is properly
discussed in the Application 1 to the Chapter II \cite{Courant}.
One has the quasi-linear PDE system with identical principle part $V^{\mu}_Q$
\begin{equation}
V^{\mu}_Q (\frac{\partial
P^{\nu}}{\partial x^{\mu} } +\Gamma^{\nu}_{\mu \lambda}P^{\lambda})=
-\frac{c}{\hbar}(\Gamma^m_{mn} \Phi_{\mu}^n(\gamma) + \frac{\partial
\Phi_{\mu}^n (\gamma)}{\partial \pi^n}) P^{\nu}P^{\mu}
\end{equation}\label{21}
for which we will build characteristics for the system of implicit solutions
for 4+4 extended variables
\begin{eqnarray}
\phi^1(x^0,x^1,x^2,x^3,P^0,P^1,P^2,P^3)=c^1;\cr
\phi^2(x^0,x^1,x^2,x^3,P^0,P^1,P^2,P^3)=c^2;\cr
\phi^3(x^0,x^1,x^2,x^3,P^0,P^1,P^2,P^3)=c^3;\cr
\phi^4(x^0,x^1,x^2,x^3,P^0,P^1,P^2,P^3)=c^4.
\end{eqnarray}\label{22}
Differentiation of $\phi^{\mu}$ in $x^{\nu}$ gives
\begin{equation}\label{23}
\frac{\partial
\phi^{\mu}}{\partial x^{\nu} }  + \frac{\partial
\phi^{\mu}}{\partial P^{\lambda}}(\frac{\partial
P^{\lambda}}{\partial x^{\nu}}+ \Gamma^{\lambda}_{\nu \mu}P^{\mu})=0.
\end{equation}
This equation being multiplied by
$\frac{\delta x^{\nu}}{\delta \tau} = V^{\nu}_Q$ gives the equation
\begin{equation}
\frac{\delta
\phi^{\mu}}{\delta \tau }=\frac{\partial
\phi^{\mu}}{\partial x^{\nu} }\frac{\delta x^{\nu}}{\delta \tau}  + \frac{\partial
\phi^{\mu}}{\partial P^{\lambda}}(\frac{\partial
P^{\lambda}}{\partial x^{\nu}}+ \Gamma^{\lambda}_{\nu \mu}P^{\mu})
\frac{\delta x^{\nu}}{\delta \tau}=0
\end{equation}\label{24}
or
\begin{eqnarray}
\frac{\partial
\phi^{\mu}}{\partial x^{\nu} } V^{\nu}_Q + \frac{\partial
\phi^{\mu}}{\partial P^{\lambda}}(\frac{\partial
P^{\lambda}}{\partial x^{\nu}}+ \Gamma^{\lambda}_{\nu \mu}P^{\mu})
V^{\nu}_Q \cr  = \frac{\partial
\phi^{\mu}}{\partial x^{\nu} } V^{\nu}_Q - \frac{\partial
\phi^{\mu}}{\partial P^{\lambda}}\frac{c}{\hbar}(\Gamma^m_{mn}
\Phi_{\mu}^n(\gamma) + \frac{\partial
\Phi_{\mu}^n (\gamma)}{\partial \pi^n}) P^{\lambda}P^{\mu}=0.
\end{eqnarray}\label{25}
Redefinition of the coefficients $C^{\nu +\lambda}:=-\frac{c}{\hbar}
(\Gamma^m_{mn} \Phi_{\mu}^n(\gamma) + \frac{\partial
\Phi_{\mu}^n (\gamma)}{\partial \pi^n}) P^{\lambda}P^{\mu}$ and variables
$x^{\nu +\lambda}:=P^{\lambda}$ gives a possibility to rewrite this
equation for any  $\phi = \phi_{\mu}$ as follows
\begin{equation}
\sum_{\kappa=1}^8 C^{\kappa} \frac{\partial
\phi}{\partial x^{\kappa}}=0.
\end{equation}\label{26}
Then one has the system of ODE's of characteristics
\begin{eqnarray}
\frac{\delta x^{\nu}}{\delta \tau}&=&V^{\nu}_Q,\cr
\frac{\delta P^{\nu}}{\delta \tau}&=&-V^{\mu}_Q
\Gamma^{\nu}_{\mu \lambda}P^{\lambda}-\frac{c}{\hbar}(\Gamma^m_{mn}
\Phi_{\mu}^n(\gamma) + \frac{\partial
\Phi_{\mu}^n (\gamma)}{\partial \pi^n}) P^{\nu}P^{\mu} \cr
\frac{d\pi^k}{d\tau} &=& \frac{c}{\hbar}\Phi_{\mu}^k P^{\mu}.
\end{eqnarray}\label{27}

The result of integration the one of the ``cross" combination
is as follows
\begin{eqnarray}
\frac{\delta x^0}{V^0_Q}=\frac{\delta P^0}{-V^{\mu}_Q
\Gamma^{0}_{\mu \lambda}P^{\lambda}+P^0L_{\mu}P^{\mu}},
\end{eqnarray}\label{28}
where $L_{\mu}= -\frac{c}{\hbar}(\Gamma^m_{mn} \Phi_{\mu}^n(\gamma) +
\frac{\partial
\Phi_{\mu}^n (\gamma)}{\partial \pi^n})$. If $L_0 L_{\alpha} < 0$ then one
gives implicit solution
\begin{eqnarray}
\frac{x^0}{a_{\alpha}x^{\alpha}} + T^0 =-\frac{2}
{\sqrt{4L_0 V^{\lambda}_Q \Gamma^0_{\lambda \alpha}P^{\alpha}+
(-V^{\lambda}_Q \Gamma^0_{\lambda 0}+L_{\alpha}P^{\alpha})^2}} \cr
\times \tanh^{-1}(\frac{2L_0P^0 +(-V^{\lambda}_Q \Gamma^0_{\lambda 0}+
L_{\alpha}P^{\alpha})}{\sqrt{4L_0 V^{\lambda}_Q \Gamma^0_{\lambda \alpha}
P^{\alpha}+(-V^{\lambda}_Q \Gamma^0_{\lambda 0}+L_{\alpha}P^{\alpha})^2}}),
\end{eqnarray}\label{29}
where $T^0$ is an integration constant. Explicit solution for energy is the kink
\begin{eqnarray}
P^0 =\frac{1}{2L_0}[V^{\lambda}_Q \Gamma^0_{\lambda 0}-L_{\alpha}P^{\alpha}-
\sqrt{4L_0 V^{\lambda}_Q \Gamma^0_{\lambda \alpha}P^{\alpha}+(-V^{\lambda}_Q
\Gamma^0_{\lambda 0}+L_{\alpha}P^{\alpha})^2} \cr \times \tanh(-\frac{x^0+
x^{\alpha}a_{\alpha}T^0}{2x^{\alpha}a_{\alpha}}{\sqrt{4L_0 V^{\lambda}_Q
\Gamma^0_{\lambda \alpha}P^{\alpha}+(-V^{\lambda}_Q \Gamma^0_{\lambda 0}+
L_{\alpha}P^{\alpha})^2}} )].
\end{eqnarray}\label{30}

This solution represent the lump of electron self-interacting through electromagnetic-like field in co-moving Lorentz reference frame. In the standard QED the self-interacting effects are treated as a polarization of the vacuum. It the present picture the lump is dynamically self-supporting system whose characteristics define by the system of four ODE's
\begin{eqnarray}
\frac{\delta P^{\nu}}{\delta \tau}&=&-V^{\mu}_Q  \Gamma^{\nu}_{\mu \lambda}
P^{\lambda}-\frac{c}{\hbar}(\Gamma^m_{mn} \Phi_{\mu}^n(\gamma) + \frac{\partial
\Phi_{\mu}^n (\gamma)}{\partial \pi^n}) P^{\nu}P^{\mu}.
\end{eqnarray}\label{31}
Self-interaction electron
is represented here as dynamical field system whose dynamical equilibrium is provided by affine gauge fields in projective Hilbert state space $CP(3)$ of spin/charge
degrees of freedom.

The standard approach to stability analysis instructs us to find the
stationary points. The stationary condition
\begin{eqnarray}
\frac{\delta P^{\lambda}}{\delta \tau}&=&0
\end{eqnarray}\label{32}
leads to the system of algebraic equations
\begin{eqnarray}\label{33}
V^{\mu}_Q  \Gamma^{\nu}_{\mu \lambda}P^{\lambda}+\frac{c}{\hbar}
(\Gamma^m_{mn} \Phi_{\mu}^n(\gamma) + \frac{\partial
\Phi_{\mu}^n (\gamma)}{\partial \pi^n}) P^{\nu}P^{\mu}=0.
\end{eqnarray}

One needs to investigate these equations for the stationary points
in the non-trivial case $P^{\mu}_0 \neq 0$. The probing solution in the vicinity of the stationary points $P^{\mu}_0 $ is as follows
\begin{eqnarray}\label{34}
P^{\mu}(\tau)=P^{\mu}_0 + p^{\mu} e^{\omega \tau}.
\end{eqnarray}
The gapless dispersion law discussed above arose in the flat Minkowski
space-time \cite{Le1}. In order to find the ``optical" dispersion law with a mass-gap and state-dependent attractor corresponding to finite mass of the electron
one should analyse the equation (\ref{33}). Then we come to the homogeneous
linear system
\begin{eqnarray}
\frac{\hbar \omega}{c}p^{\nu}+\frac{\hbar }{c}V^{\mu}_Q
\Gamma^{\nu}_{\mu \lambda}p^{\lambda}+ (\Gamma^m_{mn} \Phi_{\mu}^n(\gamma) +
\frac{\partial
\Phi_{\mu}^n (\gamma)}{\partial \pi^n}) p^{\mu}P^{\nu}_0 = 0.
\end{eqnarray}\label{35}
The determinant of this system is as follows
\begin{eqnarray}
D_1= (\frac{\hbar \omega}{c})^4+\alpha (\frac{\hbar \omega}{c})^3+
\beta (\frac{\hbar \omega}{c})^2+\gamma (\frac{\hbar \omega}{c})+\delta,
\end{eqnarray}\label{36}
with complicated coefficients $\alpha, \beta, \gamma, \delta$. If one puts
$K^{\nu}_{\lambda}=\frac{\hbar }{c}V^{\mu}_Q  \Gamma^{\nu}_{\mu \lambda}$ and
$M^{\nu}_{\mu}=(\Gamma^m_{mn} \Phi_{\mu}^n(\gamma) + \frac{\partial
\Phi_{\mu}^n (\gamma)}{\partial \pi^n}) P^{\nu}_0$ then
\begin{eqnarray}
\alpha=Tr(K^{\nu}_{\lambda})+ Tr(M^{\nu}_{\mu})
\end{eqnarray}\label{37}
and
\begin{eqnarray}
\beta=-\frac{\hbar}{c}[K^0_0(L_1P_0^1+L_2P_0^2+L_3P_0^3)+K^1_1(L_0P_0^0+L_2P_0^2+L_3P_0^3)
\cr +K^2_2(L_1P_0^1+L_0P_0^0+L_3P_0^3)+K^3_3 (L_1P_0^1+L_0P_0^0+L_2P_0^2)\cr
-K^0_1 L_0P_0^1-K^1_0 L_1P_0^0 -K^0_2 L_0P_0^2-K^2_0 L_2P_0^0
-K^0_3 L_0P_0^3-K^3_0 L_3P_0^0 \cr -K^1_2 L_1P_0^2-K^2_1 L_2P_0^1
-K^1_3 L_1P_0^3-K^3_1 L_3P_0^1 -K^2_3 L_2P_0^3-K^3_2 L_3P_0^2].
\end{eqnarray}\label{38}
Coefficients $ \gamma, \delta$ have higher order in the Newton's constant $G_N$ and they may be temporarily  discarded in approximate dispersion law.
This dispersion law may be written as follows
\begin{eqnarray}
 (\frac{\hbar \omega}{c})^2[(\frac{\hbar \omega}{c})^2+\alpha
 (\frac{\hbar \omega}{c})+\beta] =0.
\end{eqnarray}\label{39}
The trivial solution $\omega_{1,2}=0$ has already been discussed \cite{Le1}.
Two non-trivial solutions in weak gravitation field when
$\alpha^2 \gg \beta$ are given by the equations
\begin{eqnarray}
\hbar \omega_{3,4} =c \alpha \frac{-1 \pm \sqrt{1-\frac{4\beta}{\alpha^2}}}{2}
\approx c \alpha \frac{-1 \pm (1-\frac{2\beta}{\alpha^2})}{2};\cr
\hbar \omega_3=\frac{-c \beta}{\alpha}, \quad \hbar \omega_4=-
c \alpha+\frac{c \beta}{\alpha}.
\end{eqnarray}\label{40}
The  negative real part of these two roots of $\omega$ being
substituted in the probing function (\ref{34}) will define attractors and two
finite masses. The approach to the numerical analysis of these attractors is as follows.

1. Lets find initially solution of the non-linear system (\ref{33}). Its approximate solution in the vicinity of $P^{\mu}_{test} = (mc^2, 0, 0, 0) $ has been found by the method of Newton:
\begin{eqnarray}
P_0^{\mu} =P^{\mu}_{test} + \delta^{\mu}+...,
\end{eqnarray}\label{41}
where $\delta^{\mu}$ is the solution of the Newton's first approximation equations
\begin{eqnarray}
(2L_0mc+K^0_0)\delta^0+(L_1mc+K^0_1)\delta^1+\cr
(L_2mc+K^0_2)\delta^2+(L_3mc+K^0_3)\delta^3 &=& -\frac{(L_0m^2c^4+K^0_0mc^3)}{c^2} \cr
K^1_0\delta^0+(L_0mc+K^1_1)\delta^1+K^1_2\delta^2+K^1_3\delta^3 &=& -K^1_0mc \cr
K^2_0\delta^0+K^2_1\delta^1+(L_0mc+K^2_2)\delta^2+K^2_3\delta^3 &=& -K^2_0mc \cr
K^3_0\delta^0+K^3_1\delta^1+K^3_2\delta^2+(L_0mc+K^3_3)\delta^3 &=& -K^3_0mc,
\end{eqnarray}\label{42}
where $L_{\mu}=(\Gamma^m_{mn} \Phi_{\mu}^n(\gamma) + \frac{\partial
\Phi_{\mu}^n (\gamma)}{\partial \pi^n})$ are now dimensionless.

2. It has been assumed that self-interaction of charge and spin degrees of freedom comprise the energy-momentum whose distribution is encoded by field dynamics in dynamical space-time (DST) with help of two-level system represented by the qubit spinor \cite{Le1}. This DST will be associated with manifold
of coordinates in Lorentz reference frame attached to LDV during the virtual
``measurement". Technically it is as follows:

Any two infinitesimally close spinors $\eta$ and $\eta+\delta
\eta$ may be formally connected with infinitesimal $SL(2,C)$ transformations
represented by ``Lorentz spin transformations
matrix'' \cite{G}
\begin{eqnarray}\label{43}
\hat{L}=\left( \begin {array}{cc} 1-\frac{i}{2}\delta \tau ( \omega_3+ia_3 )
&-\frac{i}{2}\delta \tau ( \omega_1+ia_1 -i ( \omega_2+ia_2)) \cr
-\frac{i}{2}\delta \tau
 ( \omega_1+ia_1+i ( \omega_2+ia_2))
 &1-\frac{i}{2}\delta \tau( -\omega_3-ia_3)
\end {array} \right).
\end{eqnarray}
Then ``quantum accelerations" $a_1,a_2,a_3$ and ``quantum angular velocities"
$\omega_1,
\omega_2, \omega_3$ may be found in the linear approximation from
the equation $\delta \eta = \hat{L} \eta-\eta$, or, strictly speaking, from
its consequence - the equations for the velocities $\xi$ of $\eta$ spinor variations
\begin{eqnarray}
\hat{R}\left(
  \begin{array}{cc}
    \eta^0  \cr
    \eta^1
  \end{array}
\right) =
\frac{1}{\delta \tau}(\hat{L}-\hat{1})\left(
  \begin{array}{cc}
    \eta^0  \cr
    \eta^1
  \end{array}
\right) = \left(
  \begin{array}{cc}
    \xi^0 \cr
    \xi^1
  \end{array}
\right).
\end{eqnarray}\label{44}

Now both components $\eta^0$ and $\eta^1$
subject the affine parallel transport back to the initial GCS with velocities:
$\xi^0=\frac{\delta \eta^0}{\delta \tau}=-\Gamma \eta^0 \frac{\delta \pi}{\delta \tau}$
and
$\xi^1=\frac{\delta \eta^1}{\delta \tau}=-\Gamma \eta^1 \frac{\delta \pi}{\delta \tau}$.
If one put $\pi=e^{-i\phi} \tan(\theta/2)$ then $\frac{\delta \pi}{\delta \tau}=
\frac{\partial \pi}{\partial \theta}\frac{\delta \theta}{\delta \tau}+
\frac{\partial \pi}{\partial \phi}\frac{\delta \phi}{\delta \tau}$, where
\begin{eqnarray}\label{45}
\frac{\delta \theta}{\delta \tau}=-\omega_3\sin(\theta)-((a_2+\omega_1)\cos(\phi)+
(a_1-\omega_2)\sin(\phi))\sin(\theta/2)^2 \cr
-((a_2-\omega_1)\cos(\phi)+
(a_1+\omega_2)\sin(\phi))\cos(\theta/2)^2; \cr
\frac{\delta \phi}{\delta \tau}=a_3+(1/2)(((a_1-\omega_2)\cos(\phi)-
(a_2+\omega_1)\sin(\phi))\tan(\theta/2) \cr
-((a_1+\omega_2)\cos(\phi)-
(a_2-\omega_1)\sin(\phi))\cot(\theta/2)),
\end{eqnarray}
then one has the linear system of 6 real non-homogeneous equation
\begin{eqnarray}\label{46}
\Re(\hat{R}_{00}\eta^0+\hat{R}_{01}\eta^1)&=&-\Re(\Gamma \eta^0 \frac{\delta \pi}
{\delta \tau}), \cr
\Im(\hat{R}_{00}\eta^0+\hat{R}_{01}\eta^1)&=&-\Im(\Gamma \eta^0 \frac{\delta \pi}
{\delta \tau}), \cr
\Re(\hat{R}_{10}\eta^0+\hat{R}_{11}\eta^1)&=&-\Re(\Gamma \eta^1 \frac{\delta \pi}
{\delta \tau}),
\cr
\Im(\hat{R}_{10}\eta^0+\hat{R}_{11}\eta^1)&=&-\Im(\Gamma \eta^1 \frac{\delta \pi}
{\delta \tau}),
\cr
\frac{\delta \theta}{\delta \tau}&=&F_1, \cr
\quad \frac{\delta \phi}{\delta \tau}&=&F_2,
\end{eqnarray}
giving $\vec{a}_Q(\eta^0,\eta^1,\theta, \phi,F_1,F_2),
\vec{\omega}_Q(\eta^0,\eta^1,\theta, \phi,F_1,F_2)$ as the functions of
``measured" components of LDV $(\eta^0,\eta^1)$, the local coordinates of GCS
$(\theta, \phi)$ or complex $\pi$, and 2 real
frequencies $(F_1, F_2)$. One of them gives the coset deformation acting along some
geodesic in $CP(3)$ and the second one gives the velocity of rotation of the geodesic. Since $CP(3)$ is totally geodesic manifold \cite{KN},
each geodesic belongs to some $CP(1)$ parameterized by the single complex variable
$\pi=e^{-i\phi} \tan(\theta/2)$ used above.

If hypothesis about dynamical nature of electron mass defined by self-interacting spin/charge degrees of freedom is correct then it is very natural to assume that
\begin{eqnarray}\label{47}
F_1 = \frac{\delta \theta}{\delta \tau} =\Re(\omega_{3}) =\frac{c}{\hbar}\Re(\frac{- \beta}{ \alpha}),or \cr
F_1 = \frac{\delta \theta}{\delta \tau} =\Re(\omega_{4}) =\frac{c}{\hbar} \Re(- \alpha+\frac{ \beta}{\alpha}), and \cr
F_2 = \frac{\delta \phi}{\delta \tau} =\Im(\omega_{3}) =\frac{c}{\hbar}\Im(\frac{- \beta}{ \alpha}),or \cr
F_2 = \frac{\delta \phi}{\delta \tau} =\Im(\omega_{4}) =\frac{c}{\hbar} \Im(- \alpha+\frac{ \beta}{\alpha}).
\end{eqnarray}
Solution of complicated self-consistent problem (\ref{45}), (\ref{46}), (\ref{47}) will be reported elsewhere.
\section{Conclusion}
Solutions of the ``field-shell" quasi-liner PDE's for energy-momentum distribution of self-interacting quantum electron is shortly discussed. These equations are the consequence of the conservation law of energy-momentum vector field expressed by the affine parallel transport in $CP(3)$ agrees with Fubini-Study metric. Dynamical character of the energy-momentum distribution is defined by the attractor of the characteristic equations. An approach to the numerical analysis of these attractors has been proposed.

\end{document}